\documentclass[aps,twocolumn]{revtex4}
\usepackage{epsfig}
\usepackage{color}
\usepackage{epstopdf}

\begin{document}               

\title{Generic finite size scaling for discontinuous nonequilibrium 
phase transitions into absorbing states}

\author{M. M. de Oliveira$^{1,2}$, M. G. E. da Luz$^3$, and C. E. Fiore$^4$}
\address{
$^1$Departamento de F\'{\i}sica e Matem\'atica,
CAP, Universidade Federal de S\~ao Jo\~ao del Rei,
Ouro Branco-MG, 36420-000 Brazil, \\
$^2$ Theoretical Physics Division, School of Physics and Astronomy,
University of Manchester, Manchester, M13 9PL,
UK\\
$^3$ Departamento de F\'isica, Universidade Federal do Paran\'a, 
Curitiba-PR, 81531-980, Brazil \\
$^4$ Instituto de F\'isica, Universidade de S\~ao Paulo, 
S\~ao Paulo-SP,  05314-970, Brazil
}

\date{\today}

\begin{abstract}

Based on quasi-stationary distribution ideas, a general finite size 
scaling theory is proposed for discontinuous nonequilibrium phase 
transitions into absorbing states.
Analogously to the equilibrium case, we show that quantities such as, 
response functions, cumulants, and equal area probability distributions, 
all scale with the volume, thus allowing proper estimates for the 
thermodynamic limit.
To illustrate these results, five very distinct lattice models displaying 
nonequilibrium transitions -- to single and infinitely many 
absorbing states 
-- are investigated.
The innate difficulties in analyzing absorbing phase transitions are 
circumvented through quasi-stationary simulation methods. 
Our findings (allied to numerical studies in the literature) 
strongly point to an unifying discontinuous phase transition scaling 
behavior for equilibrium and this important class of nonequilibrium 
systems.

\end{abstract}


\maketitle


Nonequilibrium phase transition (NeqPT) into absorbing states (AS) is key 
in a wide range of phenomena as \cite{marro,odor07,henkel,hinrichsen,odor04}: 
chemical reactions, interface growth, epidemics, and population dynamics. 
Likewise, it is relevant for the emergence of spatio-temporal chaos in
different classes of problems, as experimentally verified in liquid crystal 
electroconvection \cite{take07}, driven suspensions \cite{pine}, and 
superconducting vortices \cite{okuma}.
So, much has been done on continuous NeqPT, specially addressing 
universality \cite{odor04,lubeck04,henkel,henkel-2010}.
But comparatively less attention has been payed to {\it discontinuous} 
transitions in systems with AS \cite{fiore14,oliveira2}, the case, e.g., 
in catastrophic shifts processes \cite{martin2015} (bearing important 
questions regarding the influence of diffusion and disorder in creating or 
destroying AS), heterogeneous catalysis \cite{ehsasi,zgb}, ecological 
\cite{scp,deser}, granular \cite{soto}, and replicator dynamics 
\cite{fontanari}, 
cooperative coinfection \cite{coinfection}, 
language formation \cite{naming}, and social patterns \cite{castellano}.
 
Discontinuous transitions to AS conceivably require mechanisms suppressing 
the formation of absorbing minority islands induced by fluctuations 
\cite{lubeck,grassberger}. 
Also, there are strong evidences they cannot occur in 1D if the interactions 
are short-range: the absence of boundary fields would prevent the 
stabilization of compact clusters \cite{hinrichsen00}.
In spite of these presumably universal facts, a general description of 
discontinuous NeqPT, including to identify a possible scaling behavior, 
is still lacking.  

Equilibrium first-order transitions are characterized by discontinuities 
in the order parameter $\phi$ and by thermodynamic ``densities'', whose 
susceptibilities display delta-like shapes.
In finite systems, such quantities become continuous functions of the 
control parameter $\lambda$.
However, the infinite limit still can be estimated from a finite size 
scaling theory (FSS) 
\cite{fisher,binder, binder2,binder3,lee,borgs,landau,fioreprl}, when
second derivates scale linearly with the volume 
$V = L^d$ (for $d$ the spatial dimension and $L$ the lattice size).
Also $|\lambda_V - \lambda_0|$ goes with $1/V$, with $\lambda_V$ 
($\lambda_0$) the coexistence point for a finite $V$ (in the thermodynamic 
limit).

For NeqPT to AS, precise methods like spreading simulations -- available 
for continuous transitions -- as well as a FSS framework (like the above) 
are absent in the discontinuous case.
Actually, a difficulty in its analysis is that the AS often prevent 
simulations to properly converge, precluding any scaling inference.
Even for large systems, eventually the dynamics will end up in an AS via 
a statistical fluctuation of small, but nonzero, probability.
Also, metastable states can make hard to locate or even classify transition 
points due to doubts if the observed order parameter jump is genuine.

In the present contribution we address such class of problems, presenting 
solid arguments for a common finite size scaling behavior.
Based on previous suggestions \cite{rikvold,saif,sinha,fiore14} -- and 
in the fact that equilibrium and nonequilibrium phase transitions share 
important similarities when the later display stationary (steady) states 
\cite{similarities} (see below) -- we develop a FSS for transitions into 
single and infinitely many AS by means of the quasi-stationary (QS) 
concept.
We show that, in full analogy with equilibrium, standard quantities follow 
a same $1/V$ scaling.
Five models are used to illustrate our results.


The quasi-stationary probability distribution (QSPD) idea, powerful for 
continuous NeqPT \cite{qssim}, is likewise valuable here.
In very general terms, such method has as the main purpose to evade just
the absorption process.
Formally, assume at time $t$ the microstates ($\sigma$) probability 
distribution $P(\sigma,t)$ and the survival probability $P_s(t)$, i.e., 
the probability that the system is still active.
Then, the QSPD $P_{QS}(\sigma) = \lim_{t \rightarrow \infty} 
P(\sigma,t)/P_s(t)$, describes the asymptotic properties of a finite system 
conditioned to survival \cite{dickman-vidigal02,nasell01}. 
In practice, $P_{QS}$ is calculated by effectively redistributing the flux 
from the absorbing state to the system non-absorbing subspace when the 
dynamics is sufficiently close to the absorbing condition. 
In this case, although the detailed balance is not satisfied, if the
redistribution is made compatible with the QS distribution itself 
(through a self-consistent procedure, see \cite{qssim}), then the 
global balance \cite{note0} is verified in the non-absorbing subspace
of the original problem.
Furthermore, the QS distribution becomes the stationary solution of the 
modified process \cite{dickman-vidigal02}.
Thus, typical quantities in a QS ensemble usually converge to the 
corresponding stationary ones when $L \to \infty$ \cite{dickman-vidigal02}. 

For no spatial structure problems, analytic QSPDs have been obtained 
from the master equation.
Indeed, for some discontinuous transitions, including Schl\"{o}gl 
(second) \cite{schlogol72} and ZGB \cite{zgb} models, a mean field 
calculation \cite{dickman-vidigal02,zgbqs} resulted in bimodal QSPDs. 
Nevertheless, to portray QSPDs for systems with spatial structure, one 
must rely on numerical protocols.
An efficient scheme is that in \cite{qssim}, which stores and 
gradually updates a set of configurations (compatible with the
QS ensemble) visited during the time evolution.
Whenever a transition to AS is imminent, the system is ``relocated'' 
to one of the saved configurations.
This accurately reproduces the results from the much longer procedure 
of performing averages only on samples that have not visited the 
AS at the end of their respective runs.

To construct a FSS for discontinuous NeqPTs to AS, we now observe the 
following.
First, the role of inverse flux is to turn off the system natural sink, 
thus with the absorbing becoming an `usual' phase (but with most of its 
dynamics still properly taken into account through $P(\sigma,t)$, see the 
expression for $P_{QS}$ above).
Second, certainly the resulting effective problem does not become reversible, 
but it has a weaker nonequilibrium character, presenting steady states
(the global balance, restored by the inverse flux, guarantee this later 
fact \cite{gb-ss}).
Third, to a nonequilibrium steady state we always can associate a stable
probability density \cite{stable-pd}.
Very important, such stationarity allows an extended version of the 
central limit theorem to hold true. So, the corresponding distribution 
can be described by Gaussians \cite{gaussian}.

As already mentioned, in the thermodynamic limit \cite{dickman-vidigal02} 
we can expect this resulting effective to fairly reproduce the macroscopic 
transition behavior of the original system.
Moreover, it represents a discontinuous transition between two `normal' 
phases $\pm$, bearing two scales, the order parameter $\phi = \phi_{\pm}$ 
at the transition point.
Hence, in general for a finite nonetheless reasonable large $V$, the bimodal 
probability distribution is reasonably well described by a sum of two 
Gaussians (see \cite{binder,binder2,binder3,lee}) 
$P_V(\phi) = \sum_{\omega = \pm} P_{V}^{(\omega)}(\phi)$, 
with ($\tilde{\lambda} = \lambda - \lambda_0$)
\begin{equation}
P_{V}^{(\omega)}(\phi) = \frac{\sqrt{V}}{\sqrt{2 \pi}} \,
\frac{\exp[g(V) \tilde{\lambda} \phi - g(V) (\phi-\phi_\omega)^2/
(2 \chi_\omega)]}
{[F_-(\tilde{\lambda};V) + F_+(\tilde{\lambda};V)]}.
\end{equation}
$\lambda_0$ is the control parameter value at the phase transition 
in the thermodynamic limit, the $F$'s give the normalization, and 
$g(V)$ is an increasing function of $V$.
$P_{V}(\phi)$ has the expected behavior: 
for $V \rightarrow \infty$ and $\lambda = \lambda_0$, we get the 
superposition of two $\delta$ functions centered at 
$\phi=\phi_{\pm}$.
For the extensive case $g(V) = V$
\begin{equation}
F_{\pm}(\tilde{\lambda};V) = \sqrt{\chi_{\pm}} \,
\exp\left[V \tilde{\lambda} 
\left(\phi_{\pm} +  \frac{\chi_{\pm}}{2} \tilde{\lambda}
\right)\right].
\end{equation}

Now, the pseudo-transition point $\lambda_V$ can be estimated, e.g., 
from (i) the coexisting phases equal probability condition, i.e., 
equal areas of $P_{V}^{(-)}$ and $P_{V}^{(+)}$, or yet from the 
maximum of (ii) variance 
$\chi = V(\langle\phi^2\rangle-\langle\phi\rangle^2)$, and (iii) 
moment ratio (reduced cumulant) 
$U_2 = {\langle \phi^2 \rangle}/{\langle \phi \rangle^2}$.
In first order in $\tilde{\lambda}$ \cite{note1}, both 
(i) and (ii) lead to
$\lambda_V = 
\lambda_0 - V^{-1} \, \ln[\chi_{+}/\chi_{-}]/(2 (\phi_{+}+\phi_{-}))$.
For (iii), we get $\lambda_V = \lambda_0 - V^{-1} \,
(\ln[\chi_-/\chi_+] + 2 \ln[\phi_-/\phi_+])/(2 (\phi_- - \phi_+))$.
Note $|\lambda_V - \lambda_0|$ is the same if estimated via equal 
areas or maximum of $\chi$, not differing too much if derived
by the $U_2$ maximum.
Thus, distinct measures shows that $|\lambda_V - \lambda_0| \sim 1/V$, 
the usual equilibrium scaling.

This description is illustrated by periodic square lattice models simulated
from the QS approach.
For the equal area criterion, whenever $P_{V}^{(\pm)}(\phi)$ have relevant 
overlap we consider each $P_{V}^{(\omega)}(\phi)$ occupying half of the 
corresponding $\phi$ interval.

\begin{figure}[t!]
\includegraphics[scale=0.34]{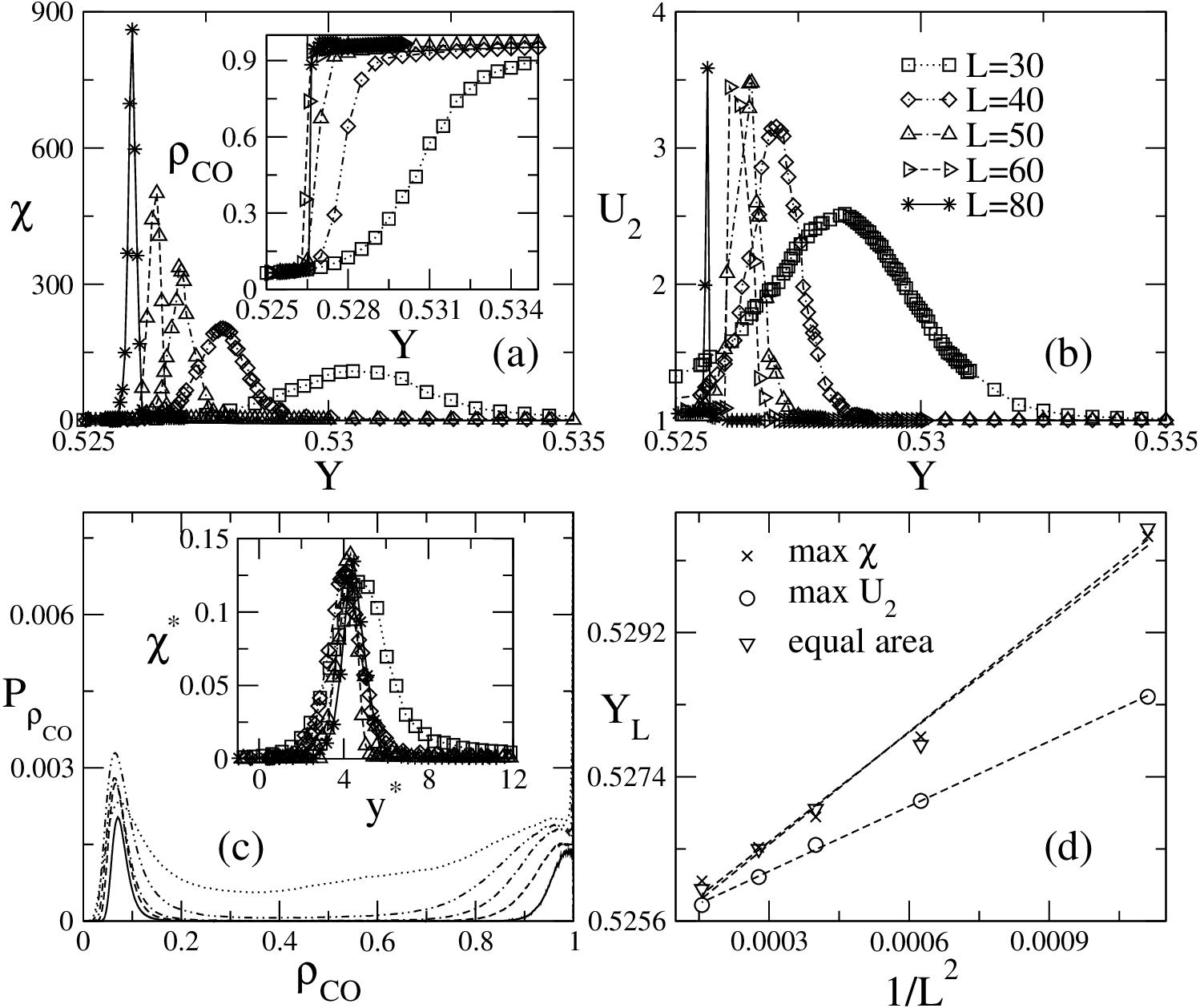}
\caption{The ZGB model.
(a) The order parameter $\rho_{CO}$ (inset) and its variance 
$\chi$ versus the creation probability $Y$. 
(b) The moment ratio $U_2$ versus $Y$.
(c) The (non-normalized) order parameter QS probability distribution 
at the equal area condition. 
Inset: data collapse analysis from the relations 
${\chi}^*=\chi/L^2$ and $y^*=(Y-Y_0)L^2$. 
(d) Scaling of $Y_L$ as function of $1/L^{2}$.}
\label{zgbfss1b}
\end{figure}

%
Consider the Ziff-Gulari-Barshad (ZGB) model \cite{zgb},
which reproduces relevant features of carbon monoxide 
oxidation on a catalytic surface (a lattice whose sites can be either 
empty or occupied by an oxygen atom O or a carbon monoxide molecule CO).
CO (O$_2$) reach the surface with probability $Y$ ($1-Y$).
Whenever a CO encounters a vacant site, the site becomes occupied.
If a O$_2$ molecule encounters two nearest-neighbor empty sites, it 
dissociates filling the two sites.
If 2 atoms O and 1 atom C reach an elementary 
$2 \times 2$ lattice cell, they immediately form CO$_{2}$ and desorb. 
The model exhibits two transitions -- regulated by the $CO$ molecules
fraction, $\rho_{CO}$ -- each between an active steady and an 
absorbing (poisoned) state.
For large (extreme low) $Y$, the surface becomes saturated 
by CO (O).
The former (latter) transition is discontinuous (continuous, 
belonging to the DP universality).
The discontinuous transition is shown in Fig. \ref{zgbfss1b}. 
The $Y$ region of rapid increase of $\rho_{CO}$ (inset of (a)) corresponds 
to the maxima of $\chi$ and $U_2$ (which increase with $L^2$, 
Fig. \ref{zgbfss1b} (a) and (b)) and their location scale with 
$1/L^{2}$, Fig. \ref{zgbfss1b} (d).
So we estimate $Y_0=0.5253(3)$ (max. of $\chi$) and $Y_0=0.5254(3)$ 
(max. of $U_2$).
The $Y_L$ for which the two peaks of $P_{\rho_{CO}}$, Fig. \ref{zgbfss1b} (c), 
have the same area also scales with $1/L^{2}$.
From this we estimate $Y_0=0.5253(3)$. 
These values are in excellent agreement among them and with 
$Y_0=0.5250(6)$, recently obtained by other means \cite{sinha}. 
Defining ${\chi}^*=\chi/L^2$ and $y^*=(Y-Y_0)L^2$, the collapsed 
data is shown in Fig. \ref{zgbfss1b} (c) inset, confirming a $L^2$ 
scaling.

\begin{figure}[t!]
\includegraphics[scale=0.34]{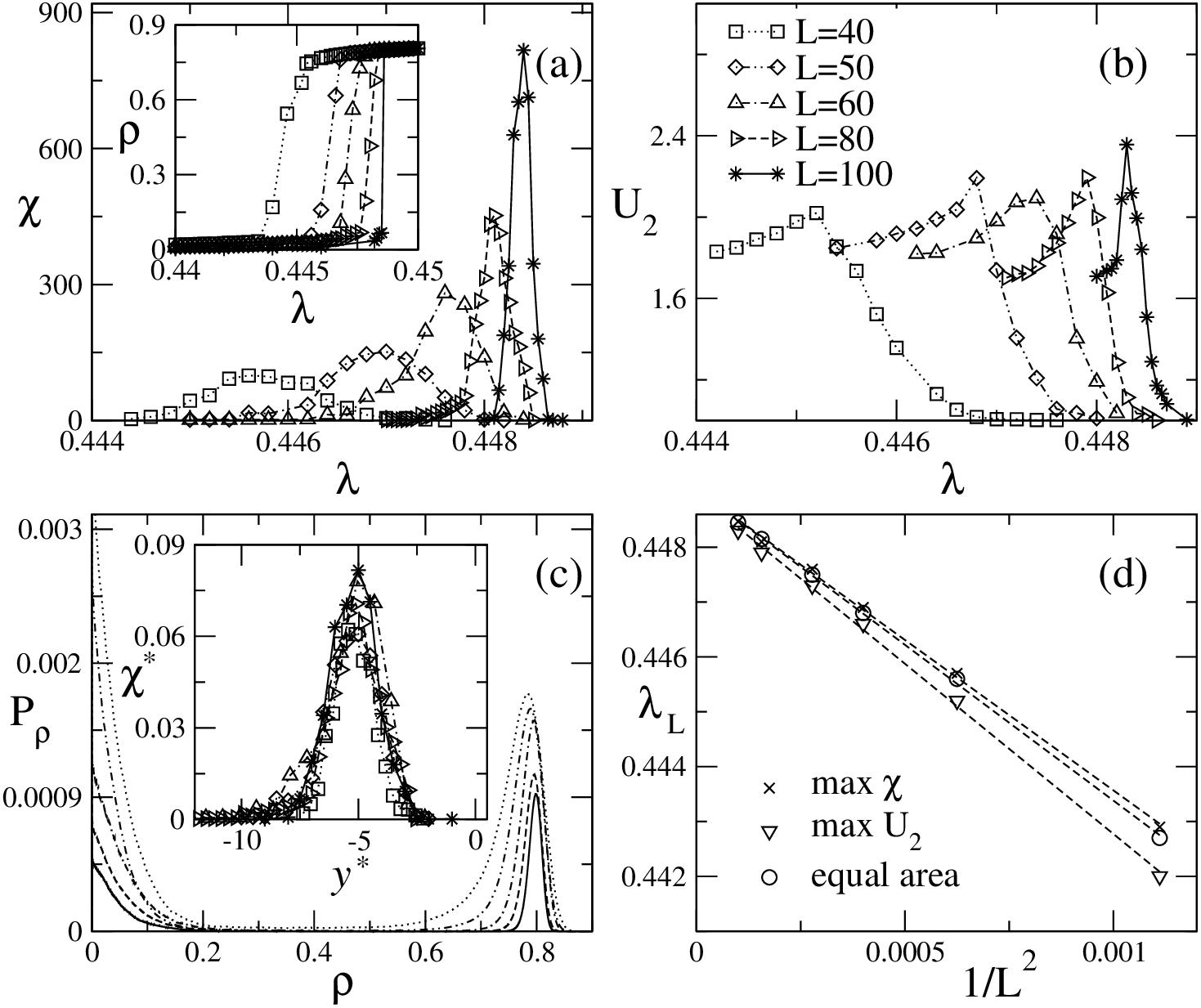}
\caption{The 2SCP model.
(a) The order parameter $\rho$ (inset) and its variance 
$\chi$ versus the creation rate $\lambda$.
(b) The moment ratio $U_2$ versus $\lambda$.
(c) The (non-normalized) order parameter QS probability distribution 
at the equal area condition. 
Inset: data collapse analysis from the relations 
${\chi}^*=\chi/L^2$ and $y^*=(\lambda-\lambda_0)L^2$. 
(d) Scaling of $\lambda_L$ as function of $1/L^{2}$.}
\label{scpfss1}
\end{figure}

%
For a two-species symbiotic contact process (2SCP) \cite{scp},
any site is either empty or occupied by an element A, by an element B, 
or by one of each. 
Each individual reproduces (autocatalytic), creating a new at 
one of its first-neighbors sites at rate 
$\lambda_A = \lambda_B = \lambda$.
In a single occupied site, A or B dies at unitary rate.
Sole individuals follows the usual CP dynamics \cite{scp}.
However, in doubly occupied sites, due to symbiosis both A and B die 
at a reduced $\mu =$ const $< 1$ rate.
Besides the CP usual active (A and B populations fixed) and 
absorbing phases, there are two extra symmetric active phases, 
in which just one species exists. 

If A and B diffuse with rate $D$, for $\mu\to 0$ 
the transition changes from continuous to discontinuous.
The order parameter is the density of occupied sites $\rho$.
Figure \ref{scpfss1} exemplifies this 2SCP for $\mu=0.01$ and $D=0.1$, 
with a discontinuous transition between absorbing and active symmetric 
phases for $\lambda \approx 0.449$ \cite{scp}.
Like ZGB, in the transition region there are peaks for 
$\chi$ and $U_2$, Fig. \ref{scpfss1} (a) and (b), whose maxima positions 
$\lambda_L$ increase with $1/L^{2}$, Fig. \ref{scpfss1} (d).
A $L \rightarrow \infty$ extrapolation yields $\lambda_0 = 0.4489(1)$ 
and $0.4490(1)$, respectively.
The equal areas condition for $P_{\rho}$, Fig. \ref{scpfss1} (c), shows a 
$1/L^{2}$ scaling, leading to $\lambda_0=0.4488(1)$. 
The estimates display excellent agreement among them and with 
Ref. \cite{scp}.
Finally, a fair data collapse is shown in Fig. \ref{scpfss1} (c) inset.

We discuss a model of competitive interactions in bipartite ($k = $ A 
and B) sublattices \cite{martins}, assuming the version in 
\cite{salete}, so instead of critical \cite{martins}, the phase diagram 
has three coexistence lines.
Also, besides an absorbing, we have a spontaneous breaking symmetry 
transition.
Given a site in the sublattice $k$, the number of particles in its first
($j=1$) and second ($j=2$) nearby neighborhood is $n_{j k}$.
For $n^{(a)}_{j k}$ the number of adjacent particles in $j$, the dynamics 
is as the following \cite{salete}.
With probability 
$(1 + \mu \, (n_{1 k})^2)/(\lambda_1 + \lambda_2 + 1 + \mu \, (n_{1 k})^2)$
we attempt to annihilate a randomly selected particle P.
If P survives, we choose at will $j=1,2$.
Then, with probability $p_j$ we try to create a new particle in 
a free site in the $j$ neighborhood of P, with 
$p_j = \lambda_j/(\lambda_1 + \lambda_2 + 1 + \mu \, (n_{1 k})^2)$ for 
$n^{(a)}_{j k} \ge j$ and zero otherwise (in \cite{martins}, 
$\mu = \lambda_2 = 0$).

\begin{figure}[t!]
\includegraphics[scale=0.34]{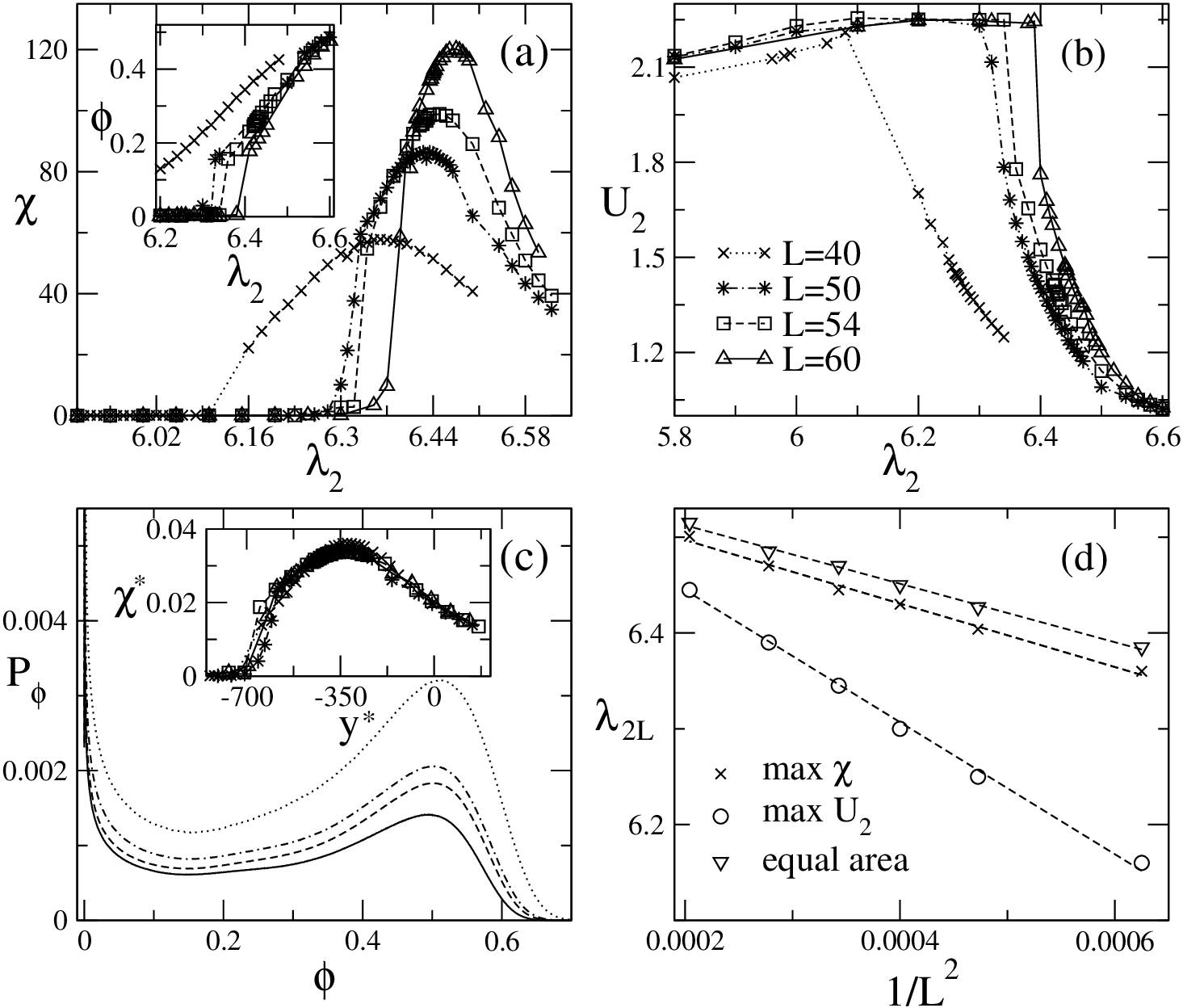}
\caption{The competitive CP ab--as transition.
(a) The order parameter $\phi$ (inset) and its variance 
$\chi$ versus the creation rate $\lambda_2$.
(a) The order parameter variance $\chi$ versus the creation rate 
$\lambda_2$.
(b) The moment ratio $U_2$ versus $\lambda_2$.
(c) The (non-normalized) order parameter QS probability distribution at 
the equal area condition. 
Inset: data collapse analysis from the relations 
${\chi}^*=\chi/L^2$ and $y^*=(\lambda_2-{\lambda_2}_0) L^2$. 
(d) Scaling of ${\lambda_2}_L$ as function of $1/L^{2}$.}
\label{fig3}
\end{figure}

\begin{figure}[t!]
\centering
\includegraphics[scale=0.32]{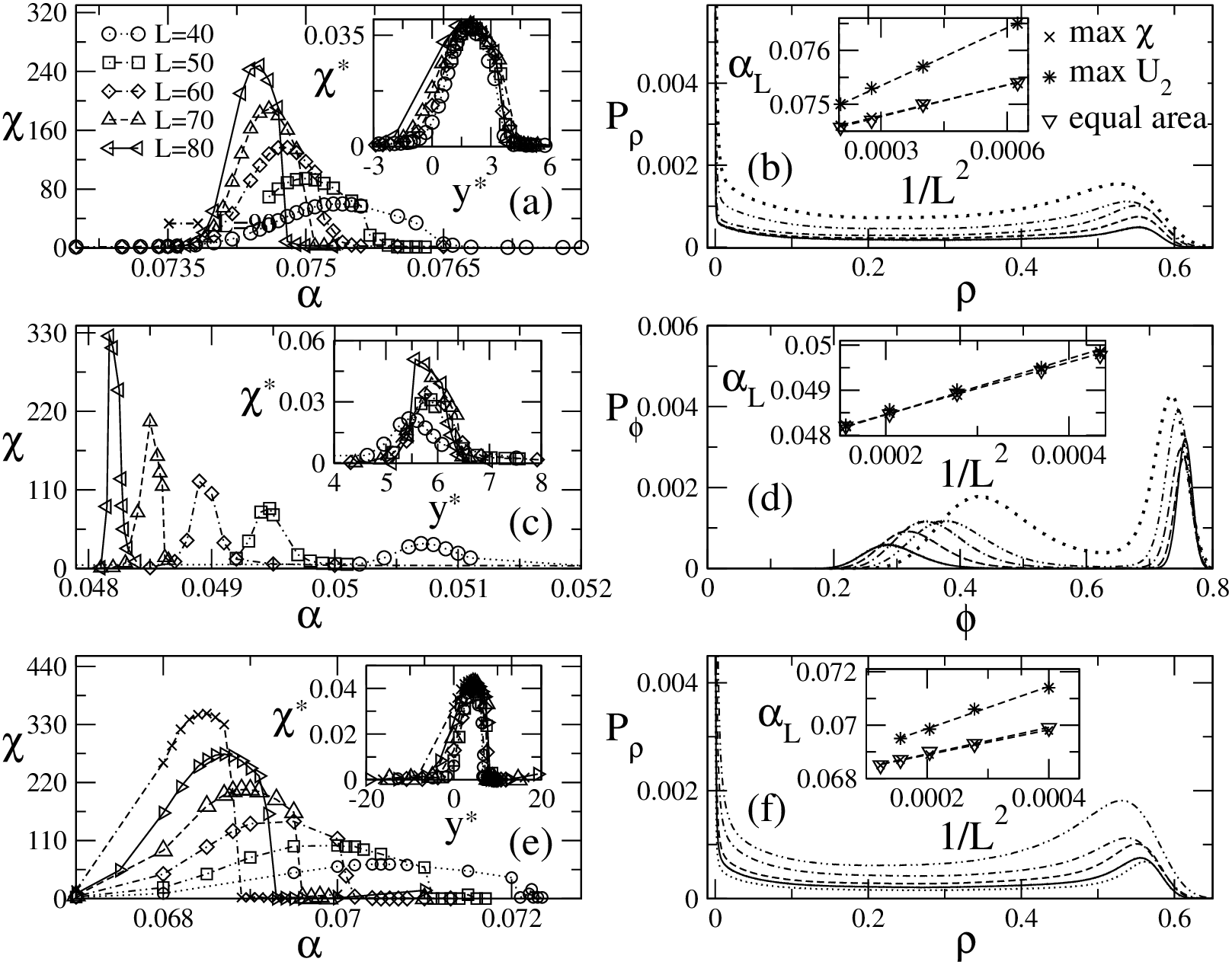}
\caption{
The second Schl\"ogl model: 
versions SL1 in (a) and (b), SL2 in (c) and (d),
and SL1 with time disorder in (e) and (f).
Left panels, the order parameters variance $\chi$ versus $\alpha$ 
(insets: their collapsed plots).
Right panels, the (non-normalized) order parameters QS probability 
distributions (insets: $\alpha_L$ as function of $1/L^2$).}
\label{fig2b}
\end{figure}

The absorbing (ab)--active symmetric (as) phases line is discontinuous for
lower  $\lambda_1$.
Proper order parameters are 
$\rho = (\rho_A+\rho_B)/2$ and $\phi=|\rho_A-\rho_B|$, with 
$\rho_X$ the $X$-sublattice density. 
In the ab phase we have $\rho = \phi = 0$, whereas for the as phase
$\rho \neq 0$ and $\phi = 0$.
So, for the as phase, the sublattices are equally populated.
From Fig. \ref{fig3} we see that the ab--as transition follows our FSS.

Finally, we address two versions of the second Schl\"ogl model 
\cite{schlogol72}: 
SL1 \cite{jensen,paula}, corresponding to a lattice version of 
the stochastic differential equation considered in \cite{martin2015}, 
and SL2 \cite{fiore14}, a modification of a pair contact process 
\cite{pcp}. 
In SL$n$, a particle ($n=1$) [a pair of two adjacent particles ($n=2$)] 
is randomly selected and can be annihilated with probability 
$p_0 = \alpha/(1+\alpha)$.
If it is not, then:\\
(1) For SL1, a nearest neighbor site $i$ is chosen.
If $i$ is empty, the particle diffuses to it.
Otherwise, with probability $p = 0.5$ \cite{jensen,paula} a new particle 
is created and placed at will in a neighboring empty site; \\
(2) If for SL2 there is at least $nn_p > 1$ other pairs in the original 
pair neighborhood, a new particle can be created with rate $nn_{p}/4$ in 
an available site in this same neighborhood.

SL1 (SL2) presents single (infinite) AS, with the order parameter being 
the particle density $\rho$ (pair density $\phi$).
The transitions occur close to $\alpha = 0.0747$ (SL1) \cite{paula} and 
$\alpha = 0.0480$ (SL2) \cite{fiore14}. 
Results are summarized in Fig. \ref{fig2b}. 
For both models our $\alpha_{L}$'s scale with $1/L^{2}$. 
For SL1, we obtain $\alpha_{0} = 0.0742(1)$ (maximum of $\chi$), 
$0.0743(1)$ (maximum of $U_2$) and $0.0742(1)$ (equal areas). 
All estimates agree very well and are close to $0.0747$ in \cite{paula}
(calculated from the threshold separating ongoing active
state and an exponential decay of $\rho$, 
considering a fully occupied initial configuration). 
For SL2 $\alpha_0 = 0.0473(1)$ (maximum of $\chi$), 
$0.0472(1)$ (maximum of $U_2$) and $0.0472(1)$ (equal areas), all close 
to $0.0480$ in \cite{fiore14} (derived from the onset for the decay 
of $\phi$ towards the absorbing regime).

Lastly, we incorporate temporal disorder into the SL1 model by assuming
that at each instance, the creation probability, $1-p_0$, is 
$\mbox{Min}\{1/(1+\alpha) + \delta,1\}$, with $\delta$ randomly chosen within 
$[- \sigma, \sigma]$. 
Results for $\sigma=0.15$ are shown in Fig. \ref{fig2b} (e) and (f). 
Here also $\alpha_{L}$'s scales with $1/L^{2}$, from which we obtain 
$\alpha_{0} = 0.0680(1)$ (maximum of $\chi$), $0.0683(2)$ (maximum of $U_2$) 
and $0.0680(1)$ (equal areas). 
Similar conclusions are obtained for $\sigma=0.25$ (not shown), from which 
$\alpha_{0} = 0.0265(1)$ (maximum of $\chi$ and equal areas).
So, in contrast to spatial disorder \cite{martin2015}, the present is 
the first evidence that temporal disorder does not hinder discontinuous 
absorbing phase transitions (but obviously, more studies should be in order,
see, e.g., \cite{vazquez-2010}).

In summary, we propose a general FSS theory for discontinuous NeqPTs to AS.
From QS ideas, we obtain an effective system -- which reproduces the 
thermodynamic properties of the original problem -- undergoing `normal' 
(i.e., not to AS) discontinuous phase transitions.
Moreover it is described by a bimodal distribution for the order parameter, 
so allowing inference of the $V$ scaling behavior.
The only eventual difficulty to implement such universal scheme would 
be if the particular system hinders a QSPD.
However, the known examples displaying such feature are very specific 
\cite{problem}.
Our study is particularly useful given that this class of NeqPTs have
no equilibrium counterparts and there are no universal treatments for 
discontinuous absorbing phase transitions for $d \ge 2$.


We acknowledge CNPq, Capes, CT-Infra and Fapesp for research grants.
We are in debt to Hans J. Herrmann and Mario de Oliveira 
for insightful discussions.

\bibliographystyle{apsrev}

\begin{thebibliography}{100}
\bibitem{note1}
Terms $O({\tilde \lambda}^2)$ are unimportant around the 
transition point for large $V$'s.
\end{thebibliography}

\end{document}